\documentclass[12pt]{article}
\usepackage{amssymb,amsmath,epsfig,cite,color}
 \allowdisplaybreaks

\begin{document}
\title{\bf Interpretation of complexity for spherically symmetric fluid composition
within the context of modified gravity theory}
\author{A. Rehman${^1}$ \thanks{v31763@umt.edu.pk}~,
Tayyab Naseer$^{2,3}$
\thanks{tayyab.naseer@math.uol.edu.pk; tayyabnaseer48@yahoo.com}~
and Baiju Dayanandan$^4$ \thanks{baiju@unizwa.edu.om}\\
${^1}$Department of Mathematics, University of Management and
Technology,\\
Johar Town Campus, Lahore-54782, Pakistan.\\
$^2$Department of Mathematics and Statistics, The University of Lahore,\\
1-KM Defence Road Lahore-54000, Pakistan.\\
$^3$Research Center of Astrophysics and Cosmology, Khazar University, \\
Baku, AZ1096, 41 Mehseti Street, Azerbaijan.\\
$^4$ Natural and Medical Sciences Research Centre,\\
University of Nizwa, Nizwa, Oman.}

\date{}

\maketitle
\begin{abstract}
Regardless of the adequate descriptions of complexity in distinct
alternative gravity theories, its elaboration in the framework of
$f(R,\mathcal{L}_{m},\mathcal{T})$ theory remains uncertain. The
orthogonal splitting of the curvature tensor yields the complexity
factor as suggested by Herrera \cite {herrera2018new}. To commence
our study, the inner spacetime is assumed to be spherically
symmetric static composition comprised of the anisotropic fluid. In
this context, we derive the modified field equations for the
considered theory and take into account the established relationship
between the conformal and curvature tensors to interpret the
complexity. Furthermore, we determine the correspondence of the mass
functions with the complexity factor, represented by a specific
scalar $Y_{TF}$. Certain solutions complying with the precedent of
diminishing $Y_{TF}$ are also evaluated. It is noted that celestial
formations having anisotropic and non-uniform compositions of matter
assert the utmost complexity. Nevertheless, the spherically
symmetric matter distribution may not exhibit complexity in the
scenario of vanishing impacts of non-homogenous energy density and
anisotropic pressure due to the presence of dark source terms
associated with this extended gravity theory.
\end{abstract}

\section{Introduction}

General relativity (GR), proposed by Albert Einstein in 1915, is
commonly regarded as the most precise description of gravity. It can
explain various gravitational events, from microscopic levels to
cosmic structures. Researchers have investigated this theory and
recognized that GR meets the standard solar system tests. At cosmic
scales, $\Lambda$-CDM model established in the framework of GR is
regarded as a particularly relevant model to understand cosmic
evolution. Later on, researchers identified some theoretical as well
as observational issues that impelled them for the generalization of
GR. Unresolved issues in GR include the existence of dark matter
(DM) at galactic as well as cosmic scales, along with the presence
of singularities inside a black hole and the mystery of dark energy
(DE). Regardless of the effectiveness of this theory in astronomical
study, it is considered to be inadequate in understanding the
expansion of the universe. Therefore, the alternative form of GR
might assist us in the detailed illustration of DM and DE.
Sufficient efforts for an appropriate gravitational framework for
the interpretation of accelerated expansion can be seen in the
literature. This endeavor is mainly specified in two main
approaches. The first emphasizes the composition of matter that
comprises most of our universe. This approach suggests that the
universe contains a mysterious negative pressure known as DE, that
causes anti-gravitational stress which eventually plays a
significant role in maintaining and accelerating the expansion of
the universe. In this regard, the cosmological constant $\Lambda$ is
added to the equations of motion associated with GR.

Nevertheless, the accelerated expansion of the universe can also be
characterized through distinct modifications of GR after considering
the specific DE frameworks. However, the cosmological constant
problem stems from this scenario which specifies the contradiction
between the theoretically evaluated value of vacuum density and the
observed value of $\Lambda$. The other technique aims to determine
the solutions for accelerating cosmic expansion through the
modification of the geometrical part. Consequently, the concept of
modified gravitational theories including multiple alternative
approaches has been emerged in the literature. Most of these
theories are specifically focused on the modified form of the linear
function of curvature invariant $R$, representing the Ricci scalar.
The formation of these alternative theories mainly relies on
generalizing the gravitational Lagrangian in the Einstein-Hilbert
action that includes a certain form $\mathcal{L}_{GR}=R$ in the
context of GR. Distinct models corresponding to these modified
theories are determined after the consideration of general curvature
functions like Ricci scalar $R$ and Gauss-Bonnet scalar $G$ along
with the impact of matter distribution that mediates from the trace
of energy-momentum tensor.

Furthermore, the consideration of modified gravity models is also
effective in the evaluation of findings related to the existence of
DE. In this regard, the $f(R)$ theory of gravity is generally
contemplated as the simplest modification of GR
\cite{buchdahl1970non,starobinsky2007disappearing}, where
$\mathcal{L}_{GR}=R$ is substituted with its general function. The
non-linear implications of the curvature scalar in the development
of the universe can be assessed through the consideration of the
appropriate form of $f(R)$ model in this particular theory of
gravity \cite{sotiriou2010f,de2010f,naseer34}. Amendola et al.
\cite{amendola2007conditions} established certain constraints
related to the feasibility of $f(R)$ DE models and evaluated their
resilience for the better description of development of the
universe. Capozziello et al. \cite{capozziello2015connecting}
studied the essential characteristics of $f(R)$ cosmology and
concluded that it is a persuasive modified form of GR for
interpreting the late-time cosmic expansion.  Nojiri and Odintsov
\cite{nojiri2006modified} suggested a general framework of DE in
this theory that can be devised after considering specific FLRW
spacetime. They evaluated plausible $f(R)$ models for defining
distinct eras of cosmic inflation.

The insertion of specific interactions between matter and geometry
yields a more extended form of $f(R)$ theories. In this regard, one
of the intriguing gravitational framework is acquired by
contemplating the Lagrangian as $f(R,\mathcal{T})$ in which
$\mathcal{T}$ specifies the trace of the energy-momentum tensor.
Bertolami along with his collaborators \cite{bertolami2007extra}
contributed significantly to establishing the connection of matter
with the structural configuration of spacetime in $f(R)$ theory of
gravity. Their technique includes the merger of matter Lagrangian
and the curvature invariant into a consolidated functional
representation termed as $f(R,\mathcal{L}_{m})$ gravity.
Subsequently, Harko et al. \cite{harko2011f} proposed a remarkable
theory, known as $f(R,\mathcal{T})$ gravity. This theory uses a
generalized function to develop a non-conserved phenomenon, leading
to the formation of an additional force that impels the motion of
particles along non-geodesic patterns \cite{deng2015solar}.
Alvarenga et al. \cite{alvarenga2013dynamics} studied the cosmic
significance of scalar disruptions associated with flat FLRW
spacetime within the context of $f(R,\mathcal{T})$ theory of
gravity. Baffou et al. \cite{baffou2017late} analyzed the late-time
development of the universe after considering the impacts of
Lagrange multipliers and mimetic potentials in this framework.
Yousaf et al. \cite{yousaf2018existence} considered the Krori-Barua
solutions in $f(R,\mathcal{T})$ theory for the analysis of the
development of anisotropic celestial objects. They claimed the
positivity of anisotropy in the relativistic structures to be
associated with the fact that the tangential component of pressure
has a stronger impact than the radial component. Bhatti et al.
\cite{bhatti2020stability,bhatti2022cylindrical,bhatti2022gravastars,
bhatti2023horizon,bhatti2023cylindrical,ur2024dynamically}
assessed the factors affecting the endurance of compact objects
having the isotropic/anisotropic fluid configuration within the
context of extended gravity theories.

It is important to mention that the inclusion of certain quantum
impacts in $f(R,\mathcal{T})$ theory might lead to the explanation
for the production of particles. This characteristic is crucial in
astronomical research because it provides the relation of modified
theory with quantum physics. The $f(R,\mathcal{T})$ theory is a
fascinating modification of GR, including several interpretations of
cosmic phenomena in the literature
\cite{yousaf1,yousaf2,yousaf3,yousaf4,t9,t10,naseer2024existence1}.
Nevertheless, researchers
\cite{velten2017cosmological,velten2021conserve} addressed the
issues of establishing a feasible cosmos within the framework of
this theory. They claimed that the recently considered models of
this theory do not properly interpret the cosmic expansion. After
considering this fact, Haghani and Harko
\cite{haghani2021generalizing} effectively unified two kinds of
gravity theories and named it the $f(R,
\mathcal{L}_{M},\mathcal{T})$ theory. This technique addresses
inadequacies in prior gravity frameworks, resulting in a more
thorough depiction of complicated cosmic interactions. They analyzed
the Newtonian limit of related equations of motion and explored the
description of accelerated expansion within the context of particles
having small velocities and weak gravitational fields. This study
also reveals the impact of distinct Lagrangian scenarios on the
interpretation of the expanding universe. Zubair et al.
\cite{zubair2023thermodynamics} reassembled the cosmic solutions
including de-Sitter and CDM frameworks, and illustrated their cosmic
stability with appropriate disruptions. Naseer and Said
\cite{naseer2024existence} investigated the feasibility of
non-singular stellar solutions under the impact of Maxwell field in
this theory. There is a large body of literature that explores
various characteristics of certain fluid configurations in different
geometric scenarios
\cite{s6,s7,s8,s9,s10,s11,z1,z4,z5,i1,i2,i3,cs2}.

The analysis of the complexity of a celestial system has been the
subject of extensive research across distinct disciplines. The
interpretation of complexity requires the contemplation of several
factors. The basic premise is associated with the measurement of
information in addition to the entropy of the fluid configuration.
The evaluation of compact objects also includes the study of
complexity related to self-gravitational systems. In the context of
physics, an impeccable crystal exhibits regular behavior and is
ordered legitimately, whereas an isolated gas exhibits chaos and
ultimate information. Both are considered complex systems with zero
intricacy. Lopez-Ruiz et al. \cite{lopez1995statistical} firstly
considered the notion of disequilibrium for the analysis of
complexity. The concept of complexity defined as the consolidation
of disequilibrium and information concluded the vanishing of
complexity in the context of both ideal gas and the perfect crystal.
Keeping this in view, the inadequacies related to the conception of
complexity, Herrera \cite{herrera2018new} suggested its new
interpretation having dependence on the distinct characteristics of
fluid including pressure and energy density. His definition is
associated with the structural features of the fluid. In this
scenario, complexity is developed through a structure scalar termed
as a complexity factor (CF) obtained from the orthogonal splitting
of the curvature tensor.

Herrera et al. \cite{herrera2018definition} extended this
description for the dissipative fluid content. The significance of
complexity in distinct structural configurations is analyzed by
Herrera and his colleagues \cite{herrera2019complexity} after
contemplating the axially symmetric composition of matter and
identified three components contributing in the complexity of the
system. The development of spherical non-static configuration within
the context of dissipative and non-dissipative fluid has also been
assessed \cite{herrera2020quasi}. They established the relationship
between areal radius velocity and areal radius within the domain of
quasi-homologous criterion and analyzed the relevance of specific
developed models in understanding the cosmic expansion. With the
help of this approach, Contreras and Fuenmayor
\cite{contreras2021gravitational} evaluated the resilience of
self-gravitational spherical compact structures via the phenomenon
of gravitating cracking. This study reviewed the density of the
object and discussed the impact of variation in decoupling factors
on pressure. Herrera et al. \cite{herrera2021hyperbolically}
illustrated the concept of complexity for hyperbolically symmetric
configuration and examined the influence of distinct structural
scalars obtained in the result of orthogonal splitting of the
curvature tensor. They derived the results for Tolman and
Misner-Sharp masses associated with this particular geometry. They
also deduced that the Tolman mass specifies negative essence in this
context. The interpretation of complexity in distinct modified
gravity theories can be seen in
\cite{yousaf2020new,naseer2024implications,naseer2024extending,sharif2024study,
arias2022anisotropic,maurya2022isotropization,maurya2023anisotropic,contreras2022uncharged,maurya2022role,
andrade2022stellar,nazar2021complexity,maurya2023complexity,bogadi2022implications,habsi2023self,
jasim2023minimally,zubair2020complexity1,zubair2020complexity,sharif2023analysis,maurya2022simple,
maurya2022relativistic,mk2024physical,ditta2024physical,naseer2024complexity,naseer2024charged,
sharif2023anisotropic,asad2022study,maurya2024compact,maurya2023complexity1,z2,z3,naseermnras,naseerpdu}.

We develop the notion of complexity within the context of $f(R,
\mathcal{L}_{m}, \mathcal{T})$ theory after getting motivation from
the work done by Herrera \cite{herrera2018new}. The main objective
of the manuscript along with the key points it addresses are
summarized as follows. The basic formation of modified theory in
addition to the fundamental concepts associated with structure and
fluid composition is discussed in section {\bf 2}. The matching
constraints related to this theory are also discussed in this
section. The external Schwarzschild structural composition is
matched with the spherically symmetric internal sector at
hypersurface $\Sigma$. Section {\bf 3} is related to the
establishment of specific relationships between active mass, Tolman
mass, and conformal tensor. The importance of all these factors in
analyzing the complexity of the system is also emphasized. The
orthogonal decomposition of intrinsic curvature results in the
illustration of structural scalars in section {\bf 4}. Out of these
scalars, one is identified as CF that plays a crucial role in
analyzing the essence of complexity. The diminishing complexity
constraint along with two specific models is studied in section {\bf
5}. Our findings are finally summarized in section {\bf 6}.

\section{Basic formulation of $f(R,\mathcal{L}_{m},\mathcal{T})$ theory}

In our case study, a mathematical framework is established in the
context of spherically symmetric dispersion of matter having
anisotropic essence of pressure. The solution for modified field
equations associated with $f(R, \mathcal{L}_{m}, \mathcal{T})$
theory are determined along with the evaluation certain kinematical
parameters that have key relevance in deriving the structure
scalars. The action related to the alternative $f(R,
\mathcal{L}_{m}, \mathcal{T})$ theory is defined after replacing the
Ricci scalar with this function \cite{naseer2024dynamic} as
\begin{equation}\label{1}
S=\int \sqrt{-g}\left[\frac{f(R,\mathcal{L}_{m},\mathcal{T})}{16 \pi}+\mathcal{L}_{m}\right]d^{4}x,
\end{equation}
where $\mathcal{L}_{m}$ specifies the matter lagrangian density.
Additionally, $g=|g_{\alpha\beta}|$ with $g_{\alpha \beta}$ being
the metric tensor and two lines encompassing it represent its
determinant. The variation of action \eqref{1} with respect to
$g_{\alpha\beta}$ yields the following form of revised equations of
motion given by
\begin{equation}\label{2}
\mathcal{G}_{\alpha \beta}= 8 \pi \mathcal{T}_{\alpha \beta}^{(eff)},
\end{equation}
where the geometrical configuration of the fluid composition is
described by $\mathcal{G}_{\alpha \beta}$ named as Einstein tensor
and the matter comprised by the contemplated geometry is denoted by
$\mathcal{T}_{\alpha \beta}^{(m)}$. The term $\mathcal{T}_{\alpha
\beta}^{(eff)}$ is precisely categorized in the following way
\begin{equation}\label{3}
\mathcal{T}_{\alpha
\beta}^{(eff)}=\frac{1}{f_{R}}\mathcal{T}_{\alpha
\beta}^{(m)}+\mathcal{T}_{\alpha \beta}^{(cr)}.
\end{equation}
Here,\\
$\bullet$ $\mathcal{T}_{\alpha \beta}^{(m)}$ relates the composition of ordinary matter.\\
$\bullet$ $\mathcal{T}_{\alpha \beta}^{(cr)}$ identifies the dark
source terms appearing due to $f(R,\mathcal{L}_{m},\mathcal{T})$
theory of gravity. Furthermore, the term $\mathcal{T}_{\alpha
\beta}^{(cr)}$ used in Eq. \eqref{3} is expressed as
\begin{eqnarray}\nonumber
\mathcal{T}_{\alpha \beta}^{(cr)}&=&\frac{1}{8 \pi
f_{R}}\left[\frac{1}{2}\left(2f_{\mathcal{T}}+f
_{\mathcal{L}_{m}}\right)\mathcal{T}_{\alpha \beta}^{(m)}-
\left(g_{\alpha \beta} \Box
-\nabla_{\alpha}\nabla_{\beta}\right)f_{R}
\right.\\\label{4}&+&\left.\frac{1}{2}\left(f-Rf_{R}\right)g_{\alpha
\beta}-\left(2f_{\mathcal{T}}+f
_{\mathcal{L}_{m}}\right)\mathcal{L}_{m}g_{\alpha \beta}+2
f_{\mathcal{T}}g^{\mu
\nu}\frac{\partial^{2}\mathcal{L}_{m}}{\partial g^{\alpha
\beta}\partial g^{\mu \nu}}\right],
\end{eqnarray}
where $f_{\mathcal{T}}=\frac{\partial f\left(R,
\mathcal{L}_{m},\mathcal{T}\right)}{\partial \mathcal{T}}$,
$f_{\mathcal{L}_{m}}=\frac{\partial f\left(R,
\mathcal{L}_{m},\mathcal{T}\right)}{\partial \mathcal{L}_{m}}$,
along with $f_{R}=\frac{\partial f\left(R,
\mathcal{L}_{m},\mathcal{T}\right)}{\partial R}$. The mathematical
forms of D'Alembertian operator and covariant derivative can be
written as $\Box \equiv
\left(-g\right)^{-1/2}\partial_{\alpha}\left(\sqrt{-g}g^{\alpha
\beta}\partial_{\beta}\right)$ and $\nabla_{\alpha}$, respectively.
This is the energy-momentum tensor corresponding to the
self-gravitational provenance which is regulated by the inner
composition of the compact object in context of
$f(R,\mathcal{L}_{M},\mathcal{T})$ theory. The energy-momentum
tensor $T^{\alpha(m)}_{\beta}$ describing the anisotropic dispersion
of fluid is represented by \cite{t1,t2,t3,t4}
\begin{equation}\label{5}
\mathcal{T}_{\beta}^{\alpha(mat)}=\rho u^{\alpha} u_{\beta}-P h_{\beta}^{\alpha}+\Pi_{\beta}^{\alpha},
\end{equation}
where
\begin{equation}\label{6}
\Pi_{\beta}^{\alpha}=\Pi\left({l^{\alpha}l_{\beta}} +\frac{1}{3}
h_{\beta}^{\alpha}\right), \quad P=\frac{P_{r}+2 P_{\bot}}{3},
\end{equation}
\begin{equation}\label{7}
\Pi=P_{r}-P_{\bot} ; \quad h_{\beta}^{\alpha}=\delta_{\beta}^{\alpha}-u^{\alpha}
u_{\beta},
\end{equation}
where $\Pi$ signifies the anisotropic pressure, while
$h_{\beta}^{\alpha}$ is called the projection tensor. The
contemplation of anisotropic pressure yields the splitting of
pressure in its radial and tangential components symbolized by
$P_{r}$ and $P_{\bot}$, respectively. Also, $\rho$ is the energy
density, $u^{\alpha}$ is the four-velocity and $l^{\alpha}$
indicates the four-vector. The internal region of the spherically
symmetric geometry is characterized through the following metric
\cite{bhatti2022analysis,t5,t6,t7,t8}
\begin{equation}\label{8}
d s^{2}=e^{\varrho(r)} d t^{2}-e^{\sigma(r)} d
r^{2}-r^{2}\left(d^{2} \theta+\sin ^{2} \theta d^{2} \phi\right).
\end{equation}
Subsequently, the expressions for four-velocity and the four-vector
are written as \textcolor{blue} {\begin{equation}\label{9}
u^{\alpha}=\left(e^{\varrho/2}, 0,0,0\right), \quad
l^{\alpha}=\left(0, e^{-\sigma / 2}, 0,0\right),
\end{equation}}
satisfying \textcolor{blue} {\begin{equation}\label{10}
l^{\alpha}l_{\alpha}=-1, \quad u^{\alpha}u_{\alpha}=1, \quad
l^{\alpha}u_{\alpha}=0.
\end{equation}}
Furthermore, the modified field equations related to $f(R, \mathcal{L}_{m}, \mathcal{T})$ gravity are
\begin{eqnarray}\label{9}
&&8\pi\left(\rho+\mathcal{T}_{0}^{0(cr)}\right)=\frac{1}{r^{2}}+\left(\frac{\sigma^{\prime}}{2}-\frac{1}{r^{2}}\right)
e^{-\sigma},
\\\label{10}
&&8\pi\left(-P_{r}+\mathcal{T}_{1}^{1(cr)}\right)=\frac{1}{r^{2}}-\left(\frac{\varrho^{\prime}}{r}+\frac{1}{r^{2}}\right)
e^{-\sigma},
\\\label{11}
&&32
\pi\left(-P_{\bot}+\mathcal{T}_{2}^{2(cr)}\right)=e^{-\lambda}\left\{\varrho^{\prime}
\sigma^{\prime}-2 \varrho^{\prime \prime}-\varrho^{\prime
2}+\frac{2\left(\sigma^{\prime}-\varrho^{\prime}\right)}{r}\right\},
\end{eqnarray}
with $\mathcal{T}_{0}^{0(cr)}$, $\mathcal{T}_{1}^{1(cr)}$ and
$\mathcal{T}_{2}^{2(cr)}$ being dark source terms and their
expressions can be seen in Appendix. The non-conserved form of
energy-momentum tensor results in the derivation of the hydrostatic
equilibrium equation in the subsequent form
\begin{eqnarray}\nonumber
&&\frac{1}{2}\varrho^{'}e^{-\sigma}\left(\rho+P_{r}\right)+\frac{1}{2}\varrho^{'}
e^{\varrho-\sigma}\mathcal{T}^{00(cr)}+\frac{1}{2}\varrho^{'}\mathcal{T}^{11(cr)}
+P_{r}^{'}e^{-\sigma}+\mathcal{T}^{11'(cr)}\\\label{12}&&+\sigma^{'}\mathcal{T}^{11(cr)}
-\frac{e^{-\sigma}}{r}P_{\bot}-re^{-\sigma}\mathcal{T}^{22(cr)}+\frac{1}{r}P_{r}e^{-\sigma}
+\frac{1}{r}\mathcal{T}^{11(cr)}=Z^{\star},
\end{eqnarray}
where the value of $Z^{\star}$ is given in Appendix. The above
equation describes the anisotropic composition of the fluid and is
termed as TOV equation, which has significant relevance in the
interpretation of the physical structure of spherically symmetric
system. Further, the usage of Eq. \eqref{10} results in
\begin{equation}\label{13}
\varrho^{'}=\frac{8 \pi r^{3}\left(P_{r}-\mathcal{T}^{11(cr)}\right)+2m}{r(r-2m)}.
\end{equation}
The non-conservation equation can be rewritten as follows after
inserting the value of $\varrho^{'}$
\begin{align}\nonumber
P_{r}^{'}&=-\left(\frac{4 \pi
r^{3}\left(P_{r}-\mathcal{T}^{11(cr)}\right)+m}{r(r-2m)}\right)\left(\rho+P_{r}
+e^{\varrho}\mathcal{T}^{00(cr)}+e^{\sigma}\mathcal{T}^{11(cr)}\right)
\\\label{14}& -e^{\sigma} \mathcal{T}^{11'(cr)}-\sigma^{'}e^{\sigma}\mathcal{T}^{11(cr)}
+r\mathcal{T}^{22(cr)}-\frac{1}{r}e^{\sigma}\mathcal{T}^{11(cr)} \times \mathcal{T}^{22(cr)}e^{\sigma}
+e^{\sigma}Z^{\star}.
\end{align}

The three-dimensional timelike hypersurface is regarded as the
precise matching between the interior and exterior sectors of
celestial objects. The matching constraints suggested by Darmois
\cite{darmois1927memorial} are considered as the most effective
conditions for the description of smooth matching between two
distinct surfaces. As an outer metric that can be matched with the
inner analog, we assume the Schwarzschild metric given by
\begin{equation}\label{15}
ds^{2}=\left(1-\frac{2 M}{r}\right)dt^{2}-\frac{dr^{2}}{\left(1-\frac{2M}{r}\right)}
-r^{2}d\theta^{2}-r^{2}\sin^{2}\theta d\phi^{2},
\end{equation}
where $M$ specifies the total mass associated with the outer
structural configuration. Furthermore, these constraints can be
described in following way

\noindent $\bullet$ Spacetimes corresponding to interior and
exterior regions must be persistent at $\Sigma$ defined by
\begin{equation}\label{D1}
[ds^{2}_{+}]_{\Sigma}=[ds^{2}]_{\Sigma}=[ds^{2}_{-}]_{\Sigma}.
\end{equation}
$\bullet$ The continuity of extrinsic curvature is holds at the
hypersurface as
\begin{equation}\label{D2}
[K_{cd}]_{\Sigma}=[K_{cd}]_{+}=[K_{cd}]_{-},
\end{equation}
whereas the mathematical form of extrinsic curvature is described as
\begin{equation}\label{D3}
K^{\pm}_{cd}=-n^{\pm}_{\beta}\left[\frac{\partial^{2}x_{\beta}}{\partial\zeta^{c}\partial\zeta^{d}}
+\Gamma^{\beta}_{\varepsilon\nu}\frac{\partial
x^{\varepsilon}}{\partial \zeta^{c}}\frac{\partial x^{\nu}}{\partial
\zeta^{d}}\right]_{\Sigma}.
\end{equation}
It is important to mention that $\zeta^{c}$ enumerates the internal
coordinates and $n^{\pm}_{\beta}$ is referred to the vector normal
to the surface. The consideration of Eqs. \eqref{D1}-\eqref{D3} at
$r=r_{\Sigma}$ concludes
\begin{equation}\label{D4}
e^{\varrho_{\Sigma}}=1-\frac{2M}{r_{\Sigma}}, \quad e^{-\sigma_{\Sigma}}=1-\frac{2M}{r_{\Sigma}},
\quad [P_{r}]_{\Sigma}=-D_{0},
\end{equation}
where the value of $D_{0}$ is presented in Appendix.

\section{The curvature tensor and its relation with other quantities}

The relation between Ricci scalar, intrinsic curvature and conformal
tensor is established through the following expression
\begin{equation}\label{7c}
R_{\varphi a  \mu}^{\gamma}=\mathcal{C}_{\varphi a
\mu}^{\gamma}+\frac{1}{2} R_{a}^{\gamma} g_{\varphi
\mu}-\frac{1}{2} R_{\varphi a} \delta_{\mu}^{\gamma}+
\frac{1}{2} R_{\varphi \mu} \delta_{a}^{\gamma} -\frac{1}{2}
R_{\mu}^{\gamma} g_{\varphi a}-\frac{1}{6}
R\left(\delta_{\mu}^{\gamma} g_{\varphi a}-g_{\varphi \mu}
\delta_{a}^{\gamma}\right).
\end{equation}
When an object crosses the geodesic, the characteristics related to
its tidal force are inferred by the conformal tensor. It can be
determined after examining the change in distance in the nearby
geodesic. The conformal tensor is categorized as the only component
of curvature tensor that addresses the propagation of gravitational
waves in the absence of matter. The trace-free essence of the
conformal tensor is considered to be its most significant
characteristic. The Riemann tensor is comprised of an electric and a
magnetic component. In the context of spherical dispersion of
matter, magnetic effects are not of any relevance because the study
of flow reveals that the behavior of line expansion is independent
of one another. Consequently, the implosion rotates around the
elements of the fluid composition. The value of the electric part in
relation with the conformal tensor can be determined by considering
the following expression
\begin{equation}\label{13}
\mathcal{E}_{\varphi \mu}=\mathcal{C}_{\varphi\gamma\mu\beta}
u^{\gamma} u^{\beta}, \quad \gamma, \beta=0,1,2,3,
\end{equation}
with
\begin{equation}\label{13}
\mathcal{C}_{\varphi\gamma\mu\beta}=\left(g_{\varphi\gamma\alpha\xi}g_{\mu\beta\sigma\nu}
-\eta_{\varphi\gamma\alpha\chi}
\eta_{\mu\beta\sigma\nu}\right)u^{\alpha}u^{\sigma}
\mathcal{E}^{\xi}\mathcal{E}^{\nu}.
\end{equation}
The alternate form is
\begin{equation}\label{13}
\mathcal{E}_{\varphi \mu}=\in\left(l_{\varphi} l_{\mu}+\frac{1}{3}
h_{\varphi \mu}\right),
\end{equation}
with $\in$ being the Weyl scalar and its value is
\begin{equation}\label{1a}
\in=\frac{-e^{-\sigma}}{4}\left[\varrho^{\prime
\prime}+\frac{\varrho^{\prime2}-\sigma^{\prime}
\varrho^{\prime}}{2}-\frac{\varrho^{\prime}-\sigma^{\prime}}{r}
+\frac{2\left(1-e^{\sigma}\right)}{r^{2}}\right],
\end{equation}
with the restrictions given below
\begin{equation}\label{13}
\in_{\varphi}^{\varphi}=0, \quad \in_{\varphi \xi}=\in_{(\varphi \xi)}, \quad
\in_{\varphi \mu} u^{\mu}=0.
\end{equation}

\subsection{Active gravitational and Tolman mass functions}

The evaluation of mass functions suggested by Misner-Sharp
\cite{misner1964relativistic} and Tolman \cite{tolman1930use} is
contemplated for the analysis of the configuration of matter. These
mathematical frameworks suggest the same outcomes at the boundary
yet imply distinct interpretations about the energy associated with
the internal region for the non-uniform dispersion of fluid. The
Misner-Sharp mass has been considered in determining the  physical
impacts of gravitational collapse
\cite{wilson1971numerical,bruenn1985stellar}, whereas the assertion
of mass proposed by Tolman \cite{bonnor2001interactions} is regarded
as plausible gravitational mass. The former mass is determined as
\begin{equation}\label{15}
m=\frac{r}{2}\left(1-e^{-\sigma}\right).
\end{equation}
The differentiation of the preceding equation in addition to the
consideration of Eq. \eqref{9} yields
\begin{equation}\label{1c}
m=\int_{0}^{r} r^{2}\left(\rho+\mathcal{T}_{0}^{0(cr)}\right) d r.
\end{equation}
On the other hand, Eqs. (\ref{9})-(\ref{11}) yields in combination
with the Misner-Sharp mass as
\begin{equation}\label{13}
m=\frac{4 \pi
r^{3}}{3}\left[\rho-2P_{\bot}-P_{r}+\mathcal{T}_{\nu}^{\mu(cr)}\right]+\frac{r^{3}}{3}
\in.
\end{equation}
The above mass function is determined in a new representation
through the insertion of Eq. \eqref{1a}. This is given as follows
\begin{eqnarray}\label{1b}
m=\frac{4 \pi
r^{3}}{3}\left[\rho-2P_{\bot}-P_{r}+\mathcal{T}_{\nu}^{\mu(cr)}\right]-\frac{r^{3}e^{-\sigma}}{12}
\left[\varrho^{\prime\prime}-\frac{\varrho^{\prime}\sigma^{\prime}}{2}+\frac{\varrho^{\prime
2}}{2}+\frac{\sigma^{\prime}}{r}-\frac{\varrho^{\prime}}{r}+\frac{2}{r^{2}}-\frac{2}{r^{2}e^{-\sigma}}\right].
\end{eqnarray}
The analogy of Eq. \eqref{1c} with \eqref{13} yields a new form of
$\in$ as follows
\begin{equation}\label{1d}
\in=4 \pi\left(P_{r}+2P_{\bot}-\mathcal{T}_{1}^{1(cr)}-2
\mathcal{T}_{2}^{2(cr)}\right) -\frac{4 \pi}{3}\int_{0}^{r}
r^{3}\left(\rho+\mathcal{T}_{0}^{0(cr)}\right)^{\prime} d r.
\end{equation}
The preceding equation characterizes the interrelation between the
conformal tensor and the properties of the spherical composition of
matter, such as non-homogenous energy density and anisotropic
pressure. Ultimately, inserting Eq. \eqref{1d} into \eqref{13}
yields
\begin{equation}\label{1e}
m=\frac{4 \pi
r^{3}}{3}\left[\rho-6P_{\bot}+\mathcal{T}_{0}^{0(cr)}-\mathcal{T}_{2}^{2(cr)}\right]-\frac{4
\pi}{3} \int_{0}^{r}
r^{3}\left(\rho+\mathcal{T}_{0}^{0(cr)}\right)^{\prime} d r.
\end{equation}
The distribution of fluid is significantly impacted by
non-homogenous energy density and anisotropic pressure through Weyl
tensor within the context of $f(R, \mathcal{L}_{m},\mathcal{T})$
theory. Whereas, in consideration of the homogenous distribution of
mass, the required deviation is enforced through non-homogeneity to
be characterized by the mass function as described in Eq.
\eqref{1e}.

Tolman \cite{tolman1930use} proposed a substantial composition for
the creation of energy associated with the matter. The Tolman mass
for the spherical configuration of matter has the following form
\begin{equation}\label{1f}
m_{T}=4 \pi \int_{0}^{r_{\Sigma}} r^{2} e^{(\varrho+\sigma) /
2}\left[ \rho+P_{r}+2 P_{\bot}\right] d r.
\end{equation}
Bhatti et al. \cite{bhatti2020stability} determined this mass
function corresponding to the spherically symmetric compositions in
the framework of alternative theory of gravity. The framework
recommended by Tolman aimed to approximate the entire mass-energy
related to the matter based on its structure. We shall demonstrate
the derivation of mass associated with the spherically symmetric
sphere with radius $r$ within the context of $f(R, \mathcal{L}_{m},
\mathcal{T})$ theory.
\begin{equation}\label{1g}
m_{T}=4 \pi \int_{0}^{r} \widetilde{r}^{2} e^{(\varrho+\sigma) /
2}\left[ \rho+P_{r}+2 P_{\bot}\right] d r,
\end{equation}
interpreting the apparent conduct of the inertial mass enumerated as
$m_{T}$, particularly elaborated in
\cite{herrera2009structure,herrera1997local}. \textcolor{blue} {The
Misner-Sharp mass accounts for the total energy (incorporating the
gravitational energy) related to the spherically symmetric matter
composition. On the other hand, the Tolman mass provides the
explanation for the gravitational mass of compact objects after the
contemplation of significant internal pressure along with the
conventional energy density estimations. It offers the better
interpretation of the relation between mass and energy associated
with a relativistic astrophysical system. Consequently, we can say
that Misner-Sharp mass is useful for measuring the entire energy
including the gravitational binding energy inside a certain radius
while the Tolman mass takes into consideration both the energy
density and the pressure components to the gravitational mass. The
combined consideration of these mass functions gives a more precise
depiction of the significant impact of pressure and mass on the
gravitational field and the structural composition of the system.}

\textcolor{blue} {The insertion of Eqs. (\ref{9})-(\ref{11}) in
\eqref{1g} yields
\begin{equation}\label{1h}
m_{T}=e^{\frac{\varrho-\sigma}{2}} \frac{\varrho^{\prime}
r^{2}}{2}+4 \pi \int_{0}^{r} r^{2}
e^\frac{\varrho+\sigma}{2}\left(\rho+P_{r}+2 P_{\bot}\right) d r.
\end{equation}
After putting the value of $\varrho^{\prime}$, preceding equation
is written as
\begin{equation}\label{1i}
m_{T}=e^{\frac{\varrho+\sigma}{2}}\left[\frac{4 \pi
r^{3}\left(P_{r}-\mathcal{T}_{1}^{1(cr)}\right)+m} {r}\right]+4 \pi
\int_{0}^{r} r^{2} e^{\frac{\varrho+\sigma}{2}}\left(\rho+P_{r}+2
P_{\bot}\right)dr,
\end{equation}
portraying the significant relevancy of $m_{T}$, referred as
effective inertial mass. Subsequently, another important expression
for the Tolman mass is
\begin{align}\nonumber
m_{T}&=M\left(\frac{r}{r_{\Sigma}}\right)^{3}-r^{3}
\int_{0}^{r_{\Sigma}} \frac{e^{\frac{\varrho+\sigma}{2}}}{r}\left[4
\pi\left(P_{\bot}-P_{r} +2\mathcal{T}_{2}^{2(cr)}+2
\mathcal{T}_{1}^{1(cr)}-\mathcal{T}_{0}^{0(cr)}\right)\right.\\\label{7q2}&\left.-\in\right]
d r+4 \pi \int ^{r_{\Sigma}}_{0}
r^{2}e^{\frac{\varrho+\sigma}{2}}\left(\mathcal{T}^{2(cr)}_{2}+\mathcal{T}^{1(cr)}_{1}-\mathcal{T}^{0(cr)}_{0}\right)dr.
\end{align}
The contemplation of Eq. \eqref{1d} implies the following form of
above equation
\begin{align}\nonumber
m_{T}&=M\left(\frac{r}{r_{\Sigma}}\right)^{3}-r^{3}
\int_{0}^{r_{\Sigma}} \frac{e^{\frac{\varrho+\sigma}{2}}}{r}\left[4
\pi\left(P_{\bot}-P_{r} +2\mathcal{T}_{2}^{2(cr)}+2
\mathcal{T}_{1}^{1(cr)}-\mathcal{T}_{0}^{0(cr)}\right)\right.\\\nonumber&\left.-4
\pi\left(P_{r}+2P_{\bot}-\mathcal{T}_{1}^{1(cr)}-2
\mathcal{T}_{2}^{2(cr)}\right) -\frac{4 \pi}{3}\int_{0}^{r}
r^{3}\left(\rho+\mathcal{T}_{0}^{0(cr)}\right)^{\prime} d r\right] d
r\\\label{7q}& +4 \pi \int ^{r_{\Sigma}}_{0}
r^{2}e^{\frac{\varrho+\sigma}{2}}\left(\mathcal{T}^{2(cr)}_{2}+\mathcal{T}^{1(cr)}_{1}-\mathcal{T}^{0(cr)}_{0}\right)dr.
\end{align}}
This equation plays an essential role in understanding the impact of
dark source terms, anisotropic pressure, and the irregular energy
density on the Tolman mass. Consequently, we assert that this
equation interprets the repercussions of non-homogenous energy
density along with the anisotropic essence of fluid configuration on
Tolman mass in the framework of $f(R, \mathcal{L}_{m}, \mathcal{T})$
theory of gravity.

\section{Orthogonal decomposition of Riemann tensor}

We consider some physical variables to understand the attributes of
anisotropic dispersion of the fluid. These variables are yielded as
the outcome of the orthogonal splitting of the Riemann tensor.
Herrera\cite{herrera2018new} evaluated the fundamental features of
the anisotropic arrangement of fluid components through the
consideration of structural scalars. These are trace and trace-free
parts of certain tensors that are relevant in the study of
considered system. In our case study, we shall employ structural
scalars to evaluate the CF related to the spherically symmetric
configuration of matter. We initiate from the tensors as follows
\cite{herrera2009structure,herrera2011role}
\begin{eqnarray}
&&Y_{a \psi}=R_{\psi\varphi a\beta}u^{\varphi}u^{\beta},
\\\label{7a}
&&Z_{a \psi}={}_{\star}R_{\psi\varphi a \beta} u^{\varphi}
u^{\beta}=\frac{1}{2} \eta_{\psi\varphi\lambda\beta} R_{a
\gamma}^{\varphi\beta} u^{\lambda} u^{\gamma},
\\\label{7b}
&&X_{a \psi }={}_{\star} R{}^{\star}_{\psi\varphi a\beta}
u^{\varphi} u^{\beta}=\frac{1}{2} \eta_{\psi\varphi}^{\lambda\beta}
R_{\lambda\beta a \gamma}^{*} u^{\varphi} u^{\gamma} \text {, }
\end{eqnarray}
in which $\star$ specifies the dual tensor while
$\eta_{\psi\varphi}^{\lambda\beta}$ signifies the Levi-Civita symbol
having values -1,1 and 0 enumerating the negative, positive and zero
fluctuation along with
$R_{\zeta\vartheta\varphi\gamma}^{\star}=\frac{1}{2}
\eta_{\lambda\beta\varphi\gamma} R_{\zeta\vartheta}^{\lambda\beta}$.
The orthogonal splitting of the Riemann tensor can be described in
the form of these tensors \cite{gomez2008kerr}. The consideration of
Eq. \eqref{7c} in addition with the revised equations of motion
yields
\begin{equation}
R_{a \gamma}^{\vartheta \varphi}=C_{a \gamma}^{\vartheta \varphi}+28 \pi
\mathcal{T}^{[\vartheta}_{[a}\delta^{\varphi]}_{\gamma]}+8 \pi
\mathcal{T}^{(eff)}\left(\frac{1}{3}\delta^{\vartheta}_{[a}\delta^{\varphi}_{\gamma]}-\delta^{[\vartheta}_{[a}\delta^{\varphi]}_{\gamma]}
\right).
\end{equation}
The insertion of Eq. \eqref{2} in preceding equation yields the
following splitting form of the curvature tensor as
\begin{equation}
R_{a \gamma}^{\vartheta \varphi}=R_{(I) a \gamma}^{\gamma
\varphi}+R_{(II) a \gamma}^{\vartheta \varphi}+R_{(III) a
\gamma}^{\vartheta \varphi}+R_{(IV) a \gamma}^{\vartheta
\varphi}+R_{(V) a \gamma}^{\vartheta \varphi},
\end{equation}
where
\begin{align}\label{S1}
R_{\text {(I)} a \gamma}^{\vartheta \varphi}&=16 \pi \rho
u^{[\vartheta} u_{[a} \delta^{\varphi]}_{\gamma]}- 16 P
h^{[\vartheta}_{[a}\delta^{\varphi]}_{\gamma]}+ 8\pi
\left(\rho-3P\right)\left(\frac{1}{3}\delta^{\vartheta}_{[a}\delta^{\varphi}_{\gamma]}-\delta^{[\vartheta}_
{[a}\delta^{\varphi]}_{\gamma]} \right),
\\\label{S2}
R_{\text {(II)} a \gamma}^{\vartheta \varphi}&=16\Pi^{[\vartheta}_{[a}\delta^{\varphi]}_{\gamma]},
\\\label{S3}
R_{\text {(III)} a \gamma}^{\vartheta \varphi}&=4 u^{[\vartheta} u_{[a} E^{\varphi]}_{\gamma]}-\eta^{\vartheta \varphi}_{\beta}\eta_{\alpha a \gamma}E^{\beta \alpha},
\\\nonumber
R_{\text {(IV)} a \gamma}^{\vartheta \varphi}&=\frac{\delta^{\varphi}_{\gamma}}{16 \pi f_{R}}\left[f_{\mathcal{T}}\mathcal{T}^{\vartheta (m)}_{a}+\frac{1}{2}f_{\mathcal{L}_{m}}\mathcal{T}^{\vartheta (m)}_{a}-\delta^{\vartheta}_{a}\Box f_{R}+\nabla ^{\vartheta}\nabla_{a}f_{R}+\frac{1}{2}\delta^{\vartheta}_{a}f\right.\\\nonumber&\left.-\frac{1}{2}\delta^{\vartheta}_{a} R f_{R}-2\delta^{\vartheta}_{a}\mathcal{L}_{m}f_{\mathcal{T}}-\delta^{\vartheta}_{a}\mathcal{L}_{m}f_{\mathcal{L}_{m}}+2 f_{\mathcal{T}}g^{\mu\nu}\frac{\partial ^{2}\mathcal{L}_{m}}{\partial \delta^{\vartheta}_{a}\partial g^{\mu \nu}}\right]-\frac{\delta^{\varphi}_{a}}{16 \pi f_{R}}\left[\mathcal{T}^{\vartheta(m)}_{\gamma}\right.\\\nonumber&\left. \times f_{\mathcal{T}}+\frac{1}{2}f_{\mathcal{L}_{m}}\mathcal{T}^{\vartheta(m)}_{\gamma}-\delta^{\vartheta}_{\gamma}\Box f_{R}+\nabla ^{\vartheta}\nabla _{\gamma}f_{R}+\frac{1}{2}\delta^{\vartheta}_{\gamma}f-\frac{1}{2}\delta^{\vartheta}_{\gamma}R f_{R}-\delta^{\vartheta}_{\gamma}\mathcal{L}_{m}f_{\mathcal{T}}\right.\\\label{S4}&\left.-\delta^{\vartheta}_{\gamma}\mathcal{L}_{m}
f_{\mathcal{L}_{m}}+2f_{\mathcal{T}}\frac{g^{\mu\nu}\partial^{2}\mathcal{L}_{m}}{\partial \delta^{\vartheta}_{\delta}\partial g^{\mu \nu}}\right],
\\\nonumber
R_{\text {(V)} a \gamma}^{\vartheta \varphi}&=\frac{\delta^{\vartheta}_{a}}{16 \pi f_{R}}\left[f_{\mathcal{T}}\mathcal{T}^{\varphi(m)}_{\gamma}+\frac{1}{2}f_{\mathcal{L}_{m}}\mathcal{T}^{\varphi(m)}_{\gamma}
-\delta^{\varphi}_{\gamma}\Box f_{R}+\nabla ^{\varphi}\nabla _{\gamma}f_{R}+\frac{1}{2}\delta^{\varphi}_{\gamma}f\right.\\\nonumber&\left.-\frac{1}{2}\delta^{\varphi}_{\gamma}R f_{R}-2\delta^{\varphi}_{\gamma}\mathcal{L}_{m}f_{\mathcal{T}}-\delta^{\varphi}_{\gamma}\mathcal{L}_{m}f_{\mathcal{L}_{m}} +2\frac{g^{\mu \nu}\partial^{2}\mathcal{L}_{m}}{\partial \delta ^{\varphi}_{\gamma}\partial g^{\mu \nu}}f_{\mathcal{T}}\right]
-\frac{\delta^{\vartheta}_{\gamma}}{16 \pi f_{R}}\left[\mathcal{T}^{\varphi(m)}_{a}\right.\\\nonumber&\left.\times f_{\mathcal{T}}+\frac{f_{\mathcal{L}_{m}}}{2}\mathcal{T}^{\varphi(m)}_{a}-\delta^{\varphi}_{a}\Box f_{R}+\nabla ^{\varphi}\nabla _{a}f_{R}+\frac{1}{2}\delta^{\varphi}_{a}f-\frac{1}{2}\delta^{\varphi}_{a}R f_{R}-2\delta^{\varphi}_{a}\mathcal{L}_{m}\right.\\\label{S5}&\left. \times f_{\mathcal{T}}-\delta^{\varphi}_{a}\mathcal{L}_{m}f_{\mathcal{L}_{m}}+\frac{2 g^{\mu\nu}\partial^{2}\mathcal{L}_{m}}{\partial \delta^{\varphi}_{a}\partial g^{\mu\nu}}f_{\mathcal{T}}\right].
\end{align}
Some other properties are defined as
\begin{equation}
\epsilon_{\gamma \delta \nu}=v^{\beta}\eta_{\beta\gamma\delta\nu},
\quad \epsilon_{\gamma \delta \nu}v^{\nu}=0,
\end{equation}
and
\begin{equation}
\epsilon^{\alpha \varphi \tau} \epsilon_{\tau \gamma\omega}=\delta_{\gamma}^{\varphi} h_{\omega}^{\alpha}-\delta_{\gamma}^{\alpha} h_{\omega}^{\varphi}+u_{\gamma}\left(u^{\alpha} \delta_{\omega}^{\varphi}-u^{\varphi} \delta_{\omega}^{\alpha}\right).
\end{equation}
In the context of a spherical configuration of matter, the conformal
curvature tensor has significant relevance in the splitting of
curvature tensor. Three specific tensors $X_{a \psi}$, $Y_{a \psi
}$, and $Z_{a \psi}$, can be determined in the form given below
based on the prior findings
\begin{eqnarray}\label{16a}
&&X_{a \psi}=\frac{8 \pi}{3} \pi \Pi_{a \psi}+\frac{2}{3}E_{a \psi
}+\frac{8 \pi}{3} \rho h_{a \psi}+\mathcal{N}_{a \psi}^{(A)},
\\\label{16b}
&&Y_{a \psi}=\frac{8 \pi}{3} \pi \Pi_{a \psi}+\frac{2}{3}E_{a
\psi}+\frac{4 \pi}{3}\left(\rho+3P\right) h_{a \psi}+\mathcal{N}_{a
\psi}^{(B)},
\\\label{16c}
&&Z_{a \psi}=\mathcal{N}_{a \psi}^{(C)}.
\end{eqnarray}
The expressions for $\mathcal{N}_{a \psi}^{(A)}$, $\mathcal{N}_{a
\psi}^{(B)}$ and $\mathcal{N}_{a \psi}^{(C)}$ are given in Appendix.
These equations specify the tensors that may help in finding
structure scalars, and the comprehensive analysis of these scalars
can be seen in \cite{herrera2009structure}. Furthermore, the trace
and trace-free parts of these scalars are written as given below and
are employed for the evaluation of several important aspects related
to the spherical composition of matter as
\begin{equation}\label{16}
X_{T}=8 \pi \rho+\mathcal{C},
\end{equation}
where the value of $\mathcal{C}$ is given in Appendix. Furthermore,
the corresponding trace-free part becomes
\begin{equation}\label{17}
X_{TF}=\frac{8 \pi \Pi}{3}-\frac{2}{3}\in.
\end{equation}
The value of $\in$ given in Eq. \eqref{1d} results the following
form of the above equation
\begin{eqnarray}\label{18}
X_{TF}=\frac{8 \pi \Pi}{3}-\frac{8
\pi}{3}\left(P_{r}+2P_{\bot}-\mathcal{T}^{1(cr)}_{1}-2\mathcal{T}^{2(cr)}_{2}\right)+\frac{8
\pi}{9}\int_{0}^{r}r^{3}\left(\rho+
\mathcal{T}^{0(cr)}_{0}\right)^{'}dr.
\end{eqnarray}
Also,
\begin{eqnarray}\label{19}
&&Y_{T}=4\pi \rho+\mathcal{D},
\end{eqnarray}
where the expression for $\mathcal{D}$ is written in Appendix. Its
corresponding trace-free part is
\begin{eqnarray}\label{20}
&&Y_{TF}=\frac{8 \pi}{3}\Pi+\frac{2}{3}\in+\mathcal{L}_{a \psi},
\end{eqnarray}
in which $\mathcal{L}_{a \psi}=\frac{N_{a \psi}}{\delta_{a}
\delta_{\psi}+\frac{1}{3} h_{a \psi}}$. The consideration of Eq.
\eqref{1d} in \eqref{20} results into
\begin{align}\label{21}
Y_{TF}&=\frac{8 \pi \Pi}{3}+\frac{8
\pi}{3}\left(P_{r}+2P_{\bot}-\mathcal{T}^{1(cr)}_{1}-2\mathcal{T}^{2(cr)}_{2}\right)+\mathcal{L}_{a
\psi}-\frac{8 \pi}{9}\int_{0}^{r}r^{3} \left(\rho +
\mathcal{T}^{0(cr)}_{0}\right)^{'}dr.
\end{align}
\textcolor{blue} {It must be noted here that all these scalars have
a dimension of square of the inverse length.} The trace-free
components mentioned in Eqs. (\ref{17}) and (\ref{20}) can be
considered for the specific description of anisotropic pressure as
given below
\begin{equation}\label{22}
X_{TF}+Y_{TF}=\frac{16 \pi \Pi}{3}+\mathcal{L}_{a \psi}.
\end{equation}
\textcolor{blue} {The orthogonal decomposition of the curvature
tensor results in the derivation of structure scalars in the context
of spherically symmetric distribution of matter. These scalars have
significant relevance in analyzing the complexity of a system as
they interpret the dispersion of energy density, anisotropic
pressure, shear, evolution and the energy transfer in a fluid.
Individually, they highlight the following characteristics of a
matter composition
\begin{itemize}
\item $Y_{T}$ is important in characterizing the entire energy
density of a system which affects the gravity interactions, mass
dispersion and the state of equilibrium.
\item $Y_{TF}$ provides the description for the non-uniform energy
density and anisotropic pressure of the system.
\item $X_{T}$ plays a crucial role in interpreting the continuous
development and mass-energy dispersion of the fluid. It specifies
the variation in the volume of a fluid with the passage of time.
\item $X_{TF}$ specifies the uneven distribution of energy
within a celestial formation.
\end{itemize}
The distribution of matter along with its dynamic changes are
important in determining the complexity of a system. The expression
for $Y_{TF}$ determines the difference between tangential and radial
components of pressure in addition with the inhomogeneous energy
density. We consider $Y_{TF}$ as the CF in the current scenario
because it determines the anisotropic and non-uniform essence of the
fluid distribution specifying the more complicated dynamics of the
system as compared to isotropic and uniform composition of matter.
The increment in these components causes the system to be more
complex.}

The relationship between $Y_{TF}$ and the active gravitational mass
is determined after putting Eq. \eqref{21} in \eqref{7q} as
\begin{align}\nonumber
m_{T}&=M\left(\frac{r}{r_{\Sigma}}\right)^{3}-r^{3}
\int_{r}^{r_{\Sigma}}
\frac{e^{\frac{\varrho+\sigma}{2}}}{r}\left[Y_{T F}-\mathcal{L}_{a
\psi}+4 \pi\left(2
\mathcal{T}_{1}^{1(D)}-2\mathcal{T}_{0}^{0(D)}\right.\right.\\\label{23}&\left.\left.+2\mathcal{T}_{2}^{2(D)}\right)\right]
d r +4 \pi \int_{r}^{r_{\Sigma}} r^{2}
e^{\frac{\varrho+\sigma}{2}}\left(
\mathcal{T}_{2}^{2(D)}+\mathcal{T}_{1}^{1(D)}-\mathcal{T}_{0}^{0(D)}\right)
d r.
\end{align}
Using Eqs. \eqref{7q2} and \eqref{23} together gives
\begin{align}\label{24}
\int_{r}^{r_{\Sigma}}\frac{e^{\frac{\varrho+\sigma}{2}}}{r}\left(\mathcal{L}_{a
\psi}-Y_{TF}\right)dr&= 4
\pi\int_{r}^{r_{\Sigma}}\frac{e^{\frac{\varrho+\sigma}{2}}}{r}\left[P_{r}-P_{\bot}+2\mathcal{T}^{1(D)}_{1}+4\mathcal{T}^{2(D)}_{2}
-2\mathcal{T}^{0(D)}_{0} +\in\right]dr.
\end{align}
Equation \eqref{24} concedes that $Y_{TF}$ interprets the
consequences of self-gravitational source of complicated
configuration relating to the anisotropic pressure along with the
non-homogenous energy density on Tolman mass within the context of
$f(R, \mathcal{L}_{m}, \mathcal{T})$ theory of gravity. Furthermore,
$Y_{TF}$ demonstrates the role of these equations in the variation
of expression for Tolman mass in contrast with the consideration of
homogenous energy dispersion and perfect fluid. On the other hand,
the expression for Tolman mass can be described as follows
\begin{eqnarray}\label{25}
m_{T}=\int_{0}^{r} r^{2}
e^{\frac{\varrho+\sigma}{2}}\left[Y_{T}-\mathcal{D}+4
\pi\left(P_{r}+2P_{\bot}+\mathcal{T}_{1}^{1(cr)}-\mathcal{T}_{0}^{0(cr)}+2
\mathcal{T}_{2}^{2(cr)}\right)\right] d r+e^{\frac{\varrho-
\sigma}{2}} \frac{\varrho^{'} r^{2}}{2}.
\end{eqnarray}
Equation \eqref{25} illustrates the direct association between
fundamental factors, Tolman mass, and dark source terms appearing
due to modified theory. It is noteworthy that $Y_{T}$ is closely
linked with the mass density. Herrera et al.
\cite{herrera2012cylindrically} evaluated the impact of $Y_{T}$ on
the Raychaudhuri equation, which is a well-known equation for the
explanation of cosmic expansion. The overhead equation entails that
Raychaudhuri equation ought to be specified after considering
$m_{T}$, regardless the framework of modified theory.

\section{Vanishing complexity}

The vanishing of CF is studied in this section through specific
models. The evaluation of complexity in distinct domains implies
that one of the structure scalars derived from the splitting of the
Riemann tensor is associated with the complexity of fluid
composition. In this study, $Y_{TF}$ is claimed as CF that
incorporates the effect of anisotropic pressure and non-uniform
energy density in addition to the dark source terms caused by the
modified theory. Five unknown variables appear in the revised
equations of motion which imply the need for two additional
restrictions in order to attain the unique solution, and the
vanishing complexity is regarded as one of these that leads to
\begin{equation}\label{26}
\Pi=\frac{1}{9}\int_{0}^{r}
r^{3}\left(\rho+\mathcal{T}^{0(cr)}_{0}\right)^{'}dr-\frac{1}{3}\left(P_{r}+2P_{\bot}-\mathcal{T}^{1(cr)}_{1}-2\mathcal{T}^{2(cr)}_{2}\right)-\frac{\mathcal{L}_{a
\psi}}{4 \pi}.
\end{equation}
This equation reveals that diminishing CF restraint suggests either
homogenous energy density along with the isotropic pressure or
non-homogenous energy density and anisotropic essence of pressure.
Furthermore, the above-mentioned equation is contemplated as
non-local equation of state (EoS) in $f(R, \mathcal{L}_{m},
\mathcal{T})$ theory of gravity. Sharif and his collaborators
observed the stability of different celestial structures by using
multiple variable EoSs \cite{s1,s2,s3,s4,s5,t11,t12}.

\subsection{Gokhroo-Mehra ansatz}

Gokhroo and Mehra \cite{gokhroo1994anisotropic} focused on the
interior configuration of spherically symmetric dispersion of fluid
to understand the dynamics of astronomical objects. In this context,
the energy density appears to be
\begin{equation}
\rho=\left(1-\frac{\mathcal{Y} r^{2}}{r_{\Sigma}^{2}}\right)
\rho_{0},
\end{equation}
where $\mathcal{Y}$ is a constant having values with the range
$(0,1)$. Its insertion in Eq. \eqref{1c} results in
\begin{equation}\label{27}
m=\frac{4 \pi}{3} r^{3} \left(1-\frac{3 \mathcal{Y} r^{2}}{5
r^{2}_{\Sigma}}\right)\rho_{0}+4 \pi \int_{0}^{r} r^{2}
\mathcal{T}_{0}^{0(cr)} d r.
\end{equation}
The combined usage of Eq. \eqref{27} and \eqref{15} yields
\begin{equation}\label{28}
e^{-\lambda}=1-8 \pi r^{2} \xi\left[1-\frac{3 \mathcal{Y} r^{2}}{5
r^{2}_{\Sigma}}\right]-\frac{8 \pi}{r} \int_{0}^{r} r^{2}
\mathcal{T}_{0}^{0(cr)} d r.
\end{equation}
Further, the consideration of second and third equations of motion
results in the following mathematical expression
\begin{eqnarray}\label{29}
8
\pi\left\{P_{r}-P_{\bot}-\mathcal{T}_{1}^{1(cr)}+\mathcal{T}_{2}^{2(cr)}\right\}=e^{-\sigma}\left[\frac{\varrho^{\prime}}{2}+\frac{1}{r^{2}}+\varrho^{\prime}
\sigma^{\prime}-2 \varrho^{\prime \prime}-\varrho^{\prime2}+\frac{2
\sigma^{\prime}}{r} -\frac{2
\varrho^{\prime}}{r}\right]-\frac{1}{r^{2}}.
\end{eqnarray}

The incorporation of new variables having the following form has
relevant significance \textcolor{blue} {\begin{equation}\label{30}
e^{\varrho(r)}=\frac{1}{e^{ \int_0^r \left( \frac{2}{r}-2
h(r)\right) d r}},  \quad  e^{\sigma  (r)}=\frac{1}{f}.
\end{equation}}
Equation \eqref{29} can be written as follows in the context of
these newly introduced variables
\begin{equation}\label{31}
f^{\prime}-f\left[\frac{6}{2 r}-\frac{5}{2r^{2} h}-\frac{2 h^{\prime}}{h}-2 h\right]=\frac{4 \pi}{h}\left[\mathcal{T}_{1}^{1(cr)}-\mathcal{T}_{2}^{2(cr)}-\Pi-\frac{1}{r^{2}}\right].
\end{equation}
This expression appears to be identical to the one proposed by
Ricatti and is established through the analysis of $\lambda$
function corresponding to the line element that is preserved in the
variable $f(r)$ and mentioned in Eq. \eqref{28} along with the value
of $\Pi$ that is determined from Eq. \eqref{26}. Consequently, the
metric can be written having dependence on $h$ and $\Pi$ after the
consideration of the integrated form of the above equation in the
context of $f(R, \mathcal{L}_{m}, \mathcal{T}$) theory of gravity
\cite{herrera2008all}. This takes the form \textcolor{blue}
{\begin{eqnarray}\nonumber &&d s^{2}=e^{2 \int_0^r
\left(h-\frac{1}{r}\right) d r} d t^{2}-r^{2}\left(d^{2} \theta+\sin
^{2} \theta d^{2} \phi\right)\\\label{z}&&-\frac{h^{2} e^{2 \int_0^r
\left(h+\frac{5}{2r^{2} h}\right) d r}}{4 \pi r^{6} \int_0^r
\frac{h}{r^{6}}\left[\frac{1}{r^{2}}+\Pi+\mathcal{T}_{2}^{2(cr)}-\mathcal{T}_{1}^{1(cr)}\right]
e^{2 \int_0^r \left(h+\frac{5}{2r^{2}h}\right) d r} d r+F}dr^2,
\end{eqnarray}}
where $F$ specifies the constant of integration. Moreover, the
fundamental parameters are stated in the following form after the
consideration of the newly introduced variables
\begin{eqnarray}\label{7w}
&&4 \pi\rho=-4\pi\mathcal{T}_{0}^{0(cr)}+\frac{m^{\prime}}{r^{2}},
\\\label{7u}&&
4 \pi P_{r}=4 \pi \mathcal{T}_{1}^{1(cr)}+\frac{h(2 m-r)+1-\frac{m}{r}}{r^{2}},
\\\label{7v}
&&8 \pi P_{\bot}=8\pi \mathcal{T}_{2}^{2(cr)}+\left(1-\frac{2
m}{r}\right)\left[4h^{\prime}+\frac{2}{r^{2}}+h^{2}-\frac{h}{r}-\frac{1}{4
r}\right]+h\left[\frac{m}{r^{2}}-\frac{m^{\prime}}{r}\right].
\end{eqnarray}
The positive energy density along with the restraint $ \rho > P_{r},
P_{\bot}$ implies that the establishment of significant approaches
to clarify the gravitating framework remains consistent. In
addition, these equations may play a significant role in
interpreting some ambiguous yet interesting features of spherically
static composition of matter. Di Prisco \cite{di2011expansion}
derived these solutions in the framework of GR. The formation of
singularities caused by the considerable parameters associated with
the hypersurface can be prevented through the contentment of Darmois
matching constraints after taking into account any suitable exterior
solution for the considered interior geometry.

\subsection{Polytropic equation of state}

The polytropic EoS has important repercussions for evaluating the
self-gravitational composition of matter. In our case study, we
contemplate the polytropic configuration in the context of the
aforementioned vanishing CF constraint. Now, we review two distinct
polytrope scenarios \cite{herrera2013newtonian,
herrera2013general,herrera2016cracking}. Subsequently, we commence
as follows
\begin{equation}\label{32}
P_{r}=\mathcal{K}\rho^{\varphi}=\mathcal{K}\rho^{1+\frac{1}{n}},
\quad \varphi=1+\frac{1}{n}.
\end{equation}
Here, $\mathcal{K}$ is the polytropic constant, while $\psi$ and $n$
represent the polytropic exponent and polytropic index,
respectively. The solution for dimensionless equation can be
determined easily, we, therefore, incorporate some extra parameters
to derive the TOV equation along with mass function in dimensionless
representation. The variables are
\begin{equation}\label{33}
\eta=\frac{P_{rb}}{\rho_{b}}, \quad r=\frac{\zeta}{C}, \quad
C^{2}=\frac{4 \pi \rho_{b}}{\eta(n+1)}, \quad
\gamma(\zeta)=\frac{C^{3}m(r)}{4 \pi \rho_{b}}, \quad
\varphi^{n}=\frac{\rho}{\rho_{b}},
\end{equation}
where subscript $b$ specifies the evaluation of the expression at
the center. We consider $\varphi(\zeta_{\Sigma})=0$ on the boundary
$r=r_{\Sigma}(\zeta=\zeta_{\Sigma})$. The TOV equation in the
context of these variables turns out to be
\begin{eqnarray}\nonumber
&&2\zeta^{2} \frac{d \varphi}{d \zeta}\left[\frac{1-2 \eta(n+1)
\gamma / \zeta}{1+\eta
\varphi+\frac{\mathcal{T}_{1}^{1(cr)}-\mathcal{T}_{0}^{0(cr)}}{\varphi^{n}\rho_{b}}}\right]+16
\pi \eta+8 \pi r \zeta^{3} \varphi^{1+n} +\frac{8 \pi
\zeta^{3}}{\rho_{b}} \mathcal{T}_{1}^{1(D)}\\\label{SAAA}&&+\frac{2
\zeta}{P_{r
b}(n+1)}\left[\left(1-\frac{2\eta(n+1)\gamma}{\zeta}\right)\left(\frac{\Pi
+\mathcal{T}^{1(cr)}_{1}-\mathcal{T}^{2(cr)}_{2}+Z^{\star}}{\varphi^{n}\left(1+\eta
\varphi
+\frac{\mathcal{T}^{1(cr)}_{1}+\mathcal{T}^{0(cr)}_{0}}{\varphi^{n}\rho_{b}}\right)}\right)\right]=0.
\end{eqnarray}
Equation \eqref{1c} takes the following form after the contemplation
of variables \eqref{33} as
\begin{equation}\label{AAAA}
\frac{d \gamma}{d \zeta}=2 \pi
\zeta^{2}\left(\varphi^{n}+\frac{\mathcal{T}^{0(cr)}_{0}}{\rho_{b}}\right).
\end{equation}
The presence of $\Pi$, $\gamma$ and $\varphi$ functions in the
preceding two equations implies the need for one more restraints for
the derivation of precise solution for our system. In this context,
we assume the vanishing CF restraint in the form of dimensionless
variables as given below
\begin{equation}\label{SAA}
\frac{\zeta}{n \rho_{b}} \frac{d \Pi}{d \zeta}=\frac{\zeta^{4}}{9 Cn
\rho_{b}} \varphi^{n-1} \frac{d \varphi}{d \zeta}+\frac{\zeta}{n
\rho_{b}} \frac{d}{d \zeta}\left(\mathcal{T}_{0}^{0(cr)}-\chi
\right).
\end{equation}
where the value of $\chi$ can be seen in Appendix. Currently, there
are three differential equations corresponding to three unknown
functions $\Pi$, $\eta$, and $\psi$. The analytical integrated form
of these equations might be associated with different values of
$\gamma$ and $n$, or their numerical solution may be determined
after the contemplation of appropriate conditions. Every solution
could determine the pressure, mass and density for cosmic
compositions based on the specific values relating to the free
variables.

Furthermore, the second polytropic EoS can be considered as
\begin{equation}
P_{r}=\mathcal{K}\rho^{\varphi}_{d}=\mathcal{K}\rho^{1+\frac{1}{n}}_{d},
\quad \varphi=1+\frac{1}{n},
\end{equation}
where $\rho_{d}$ being the baryonic mass density. The combined usage
of this equation with \eqref{SAAA} and \eqref{SAA} yields
\begin{eqnarray}\nonumber
&&2\zeta^{2} \frac{d \varphi_{d}}{d \zeta}\left[\frac{1-2 \eta(n+1)
\gamma / \zeta}{1+\eta
\varphi_{d}+\frac{\mathcal{T}_{1}^{1(cr)}-\mathcal{T}_{0}^{0(cr)}}{\varphi^{n}\rho_{bd}}}\right]+16
\pi \eta+8 \pi r \zeta^{3} \varphi^{1+n}_{d} +\frac{8 \pi
\zeta^{3}}{\rho_{bd}} \mathcal{T}_{1}^{1(D)}\\\label{L1}&&+\frac{2
\zeta}{P_{r
b}(n+1)}\left[\left(1-\frac{2\eta(n+1)\gamma}{\zeta}\right)\left(\frac{\Pi
+\mathcal{T}^{1(cr)}_{1}-\mathcal{T}^{2(cr)}_{2}+Z^{\star}}{\varphi^{n}_{d}\left(1+\eta
\varphi_{d}
+\frac{\mathcal{T}^{1(cr)}_{1}+\mathcal{T}^{0(cr)}_{0}}{\varphi^{n}_{d}\rho_{bd}}\right)}\right)\right]=0,
\end{eqnarray}
and
\begin{equation}\label{L2}
\frac{\zeta}{n \rho_{bd}} \frac{d \Pi}{d \zeta}=\frac{\zeta^{4}}{9
Cn \rho_{bd}} \varphi^{n-1}_{d} \frac{d \varphi_{d}}{d
\zeta}+\frac{\zeta}{n \rho_{bd}} \frac{d}{d
\zeta}\left(\mathcal{T}_{0}^{0(cr)}-\chi \right),
\end{equation}
in which $\psi_{d}^{n}=\frac{\rho_{d}}{\rho_{bd}}$. Subsequently,
through the solution of Eqs. \eqref{L1}, \eqref{L2} and \eqref{AAAA}
after the contemplation of EoS \eqref{32} in addition with the
vanishing complexity restraint, the evolution of celestial
compositions can be analyzed. Equation \eqref{32} along with radial
pressure and the energy density has significant relevance in
interpreting different eras of cosmic expansion depending on the
values of $\mathcal{K}$. For $\mathcal{K}=0, \frac{1}{3}$, and $1$
it specifies the dominance of matter, radiation and the stiff fluid,
respectively. Also, the phantom phase can be evaluated by
considering $\mathcal{K} < -1$ and $\mathcal{K}\in (\frac{-1}{3},
-1)$ signifies the quintessence era. Moreover, different structural
compositions can explained by distinct values of $n$, i.e., the
values between $n=0.5$ and $n=1$ are most suitable choices for
characterizing neutron stars.

\section{Discussion and Conclusions}

Our manuscript mainly focused on interpreting Herrera's specific
approach to defining complexity within the framework of $f(R,
\mathcal{L}_{m}, \mathcal{T}$) theory. This gravitational theory
provides a plausible evaluation of the dynamics related to the
accelerating cosmic expansion. Our analysis is initiated with the
fundamental formalism of the considered theory along with some
essential parameters. The modified equations of motion have then
been determined in the context of the spherical anisotropic
distribution of matter. To further proceed, TOV equation is
determined by considering the law of non-conservation. The certain
matching constraints for the smooth matching of the external sector
with the inner configuration of spherically symmetric matter are
also derived within the modified framework. \textcolor{blue} {We
have also employed the expressions for Misner-Sharp and Tolman mass
functions which eventually play a significant role in establishing
the relationship between fundamental parameters and the conformal
tensor.} Furthermore, the impact of former mass, conformal tensor
and the correction terms on the composition of matter associated
with the anisotropic pressure and non-uniform energy density has
been evaluated. The orthogonal decomposition of the curvature tensor
yielded the expression for CF, i.e., $Y_{TF}$. Structure scalars are
obtained through this approach and we established the restraint of
the vanishing complexity. The dynamics of varying energy density
suggested by Gokhroo and Mehra \cite{gokhroo1994anisotropic} has
been analyzed in the context of diminishing complexity. Eventually,
the polytropic EoS is considered in addition to the diminishing
complexity for the derivation of viable solutions related to the
spherically symmetric configuration. \textcolor{blue} {We conclude
our analysis with the following remarks
\begin{itemize}
\item Several variables specifically anisotropic distribution of
matter and non-uniform energy density have significant relevance in
determining the complexity of a system. The structural parameters
suggested by Herrera \cite{herrera2009structure} are derived through
the orthogonal decomposition of the Riemann tensor. We evaluated
four structure scalars and studied their associated characteristics.
\item The scalar $Y_{TF}$ incorporates the effect of non-uniform energy
density, correction terms associated with $f(R, \mathcal{L}_{m},
\mathcal{T}$) theory, and anisotropic configuration of matter on the
overall energy distribution of the system.
\item This conception has also been implemented for the dissipated and
non-dissipated fluid configuration \cite{herrera2011role}. The
primary contenders for understanding the evolution of the shear and
expansion tensors are $Y_{T}$ and $Y_{TF}$.
\item The compatibility of vanishing complexity restraint is sustained
through the consideration of specific models to assess the system.
The ``Gokhroo and Mehra model'' establishes the interpretations for
fundamental parameters in the framework of considered $f(R,
\mathcal{L}_{m}, \mathcal{T})$ theory. Two unknown functions are
employed in this regard. and specific expressions for fundamental
parameters are determined in the scheme of generating functions
$\Pi$ and $h$.
\item Further, after the consideration of polytropic EoS, the expressions
for the TOV equation, vanishing CF, and mass are determined. It is
significantly important to mention that CF vanishes in the context
of isotropic pressure along with the homogenous energy density in
the framework of GR \cite{herrera2018new}.
\end{itemize}}

\noindent Regardless of these restraints, our results are consistent
with the zero complexity which demonstrates the relevance of the
dark source terms caused by $f(R, \mathcal{L}_{m}, \mathcal{T})$
theory. These results could be used to precisely interpret important
astronomical happenings such as supernova explosions. Certain
significant models ought to be established for the comprehensive
study of the relationship between $f(R, \mathcal{L}_{m},
\mathcal{T}$) theory and the interpretation of the expanding cosmos.

\section*{Appendix}

\renewcommand{\theequation}{A\arabic{equation}}
\setcounter{equation}{0} \noindent The modified corrections
associated with Eqs. \eqref{9}-\eqref{11} are
\begin{align}\nonumber
\mathcal{T}^{00(cr)}&=\frac{f_{\mathcal{T}}e^{\varrho}\rho}{8 \pi
f_{R}}+\frac{f_{\mathcal{L}_{m}}e^{\varrho}\rho}{16 \pi
f_{R}}+\frac{\varrho^{'}e^{\varrho-\sigma}}{16 \pi
f_{R}}f_{R}^{'}+\frac{e^{\varrho-\sigma}}{8 \pi
f_{R}}f_{R}^{''}-\frac{\sigma^{'}e^{\varrho-\sigma}}{16 \pi
f_{R}}f_{R}^{'}+\frac{e^{\varrho-\sigma}}{4 \pi
f_{R}}f_{R}^{'}-\frac{\varrho^{'}e^{\varrho-\sigma}}{16 \pi
f_{R}}\\\nonumber& \times f_{R}^{'}+\frac{f e^{\varrho}}{16 \pi
f_{R}}-\frac{R f_{R}}{16 \pi
f_{R}}e^{\varrho}-\frac{f_{\mathcal{T}}\mathcal{L}_{m}e^{\varrho}}{4
\pi f_{R}}-\frac{f_{\mathcal{L}_{m}}\mathcal{L}_{m}e^{\varrho}}{8
\pi f_{R}},
\\\nonumber
\mathcal{T}^{11(cr)}&=\frac{e^{\sigma}P_{r}f_{\mathcal{T}}}{8 \pi
f_{R}}+\frac{e^{\sigma}P_{r}f_{\mathcal{L}_{m}}}{16 \pi
f_{R}}-\frac{\varrho ^{'}}{16 \pi
f_{R}}f_{R}^{'}-\frac{f_{R}^{''}}{8 \pi f_{R}}+\frac{\sigma^{'}}{16
\pi f_{R}}f_{R}^{'}-\frac{f_{R}^{'}}{4 \pi r
f_{R}}+\frac{f_{R}^{''}}{8 \pi
f_{R}}\\\nonumber&-\frac{\sigma^{'}f_{R}^{'}}{16 \pi f_{R}}-\frac{f
e^{\sigma}}{16 \pi f_{R}}+\frac{R f_{R}e^{\sigma}}{16 \pi
f_{R}}+\frac{e^{\sigma}f_{\mathcal{T}}\mathcal{L}_{m}}{4 \pi
f_{R}}+\frac{e^{\sigma}f_{\mathcal{L}_{m}}\mathcal{L}_{m}}{8 \pi
f_{R}},
\\\nonumber
\mathcal{T}^{22(cr)}&=\frac{f_{\mathcal{T}}r^{2}P_{\bot}}{8 \pi
f_{R}}+\frac{f_{\mathcal{L}_{m}}r^{2}P_{\bot}}{16 \pi f_{R}}
-\frac{\varrho^{'}r^{2}e^{-\sigma}}{16 \pi
f_{R}}f_{R}^{'}-\frac{r^{2}e^{-\sigma}}{8 \pi
f_{R}}f_{R}^{''}+\frac{r^{2}e^{-\sigma}\sigma^{'}}{16 \pi
f_{R}}f_{R}^{'}-\frac{e^{-\sigma}r}{4 \pi f_{R}}f_{R}^{'}
\\\nonumber& +\frac{r e^{-\sigma}f_{R}^{'}}{8 \pi f_{R}}-\frac{r^{2}f}{16 \pi f_{R}}+\frac{r^{2}R f_{R}}{16 \pi f_{R}}+\frac{r^{2}\mathcal{L}_{m}f_{\mathcal{T}}}{4 \pi f_{R}}+\frac{r^{2}\mathcal{L}_{m}f_{\mathcal{L}_{m}}}{8 \pi f_{R}}.
\end{align}

\noindent The value of $Z^{\star}$ is
\begin{align}\nonumber
Z^{\star}&=\frac{1}{8 \pi
+f_{\mathcal{T}}+\frac{1}{2}f_{\mathcal{L}}}\left[\frac{1}{6}\left(2
P_{r}f_{\mathcal{T}}^{'}+P_{r}f_{\mathcal{L}}^{'}+4
P_{\bot}f_{\mathcal{T}}^{'}+2
P_{\bot}f_{\mathcal{L}}^{'}+2f_{\mathcal{T}}P_{r}^{'}+4
f_{\mathcal{T}}P_{\bot}^{'}\right.\right.\\\nonumber&\left.\left.+f_{\mathcal{L}}P_{r}^{'}+2f_{\mathcal{L}}P_{\bot}^{'}\right)
+e^{2 \sigma}P_{r}f_{\mathcal{T}}^{'}+\frac{e^{2
\sigma}}{2}P_{r}f_{\mathcal{L}}^{'}-\frac{f_{\mathcal{T}}\rho^{'}}{2}+\frac{f_{\mathcal{T}}P_{r}^{'}}{2}+f_{\mathcal{T}}P_{\bot}^{'}-\frac{f_{\mathcal{L}}P_{r}^{'}}{6}
-\frac{P_{\bot}^{'}}{3}\right].
\end{align}

\noindent The values of correction terms associated with Eqs.
\eqref{16a}-\eqref{20} are
\begin{align}\nonumber
\mathcal{N}_{a \psi}^{(A)}&=\left[\frac{f_{\mathcal{T}}}{32 \pi
f_{R}}\mathcal{T}^{\alpha(m)}_{\beta}+\frac{f_{\mathcal{L}_{m}}}{64
\pi f_{R}}\mathcal{T}^{\alpha(m)}_{\beta}-\frac{\Box f_{R}}{64 \pi
f_{R}}+ \frac{f}{128 \pi f_{R}}-\frac{R f_{R}}{128 \pi
f_{R}}-\frac{f_{\mathcal{T}}\mathcal{L}_{m}}{64 \pi
f_{R}}\right.\\\nonumber&\left.-\frac{f_{\mathcal{L}_{m}}\mathcal{L}_{m}}{64
\pi f_{R}} \right]\varepsilon^{\alpha \delta}_{a}\varepsilon_{\alpha
\delta \psi}+\frac{\nabla ^{\alpha}\nabla _{\beta}f_{R}}{64 \pi
f_{R}}\varepsilon ^{\beta \delta}_{a}\varepsilon_{\alpha \delta
\psi}+\left[\frac{R}{128 \pi f_{R}}-\frac{f_{\mathcal{T}}}{64 \pi
f_{R}}\mathcal{T}^{\alpha
(m)}_{\delta}-\frac{f_{\mathcal{L}_{m}}}{128 \pi
f_{R}}\mathcal{T}^{\alpha
}_{\delta}\right.\\\nonumber&\left.+\frac{\Box f_{R}}{64 \pi
f_{R}}-\frac{\nabla ^{\alpha}\nabla _{\delta}f_{R}}{64 \pi
f_{R}}-\frac{f}{128 \pi
f_{R}}+\frac{f_{\mathcal{T}}\mathcal{L}_{m}}{32 \pi
f_{R}}+\frac{f_{\mathcal{L}_{m}}}{64 \pi }
\frac{\mathcal{L}_{m}}{f_{R}}-\frac{f_{\mathcal{T}}g^{\mu \nu}}{32
\pi f_{R}}\frac{\partial ^{2}\mathcal{L}_{m}}{\partial
g^{\alpha}_{\delta}\partial g_{\mu \nu}}\right]\varepsilon^{\beta
\delta}_{a}\varepsilon_{\alpha \beta
\psi}\\\nonumber&+\left[-\frac{f_{\mathcal{T}}}{64 \pi
f_{R}}\mathcal{T}^{\gamma(m)}_{\beta}-\frac{f_{\mathcal{L}_{m}}}{128
\pi
f_{R}}\mathcal{T}^{\gamma(m)}_{\beta}-\frac{\nabla^{\gamma}\nabla_{\beta}f_{R}}{64
\pi f_{R}} -\frac{f_{\mathcal{T}}g^{\mu \nu}}{32 \pi
f_{R}}\frac{\partial ^{2}\mathcal{L}_{m}}{\partial
g^{\gamma}_{\beta}\partial g_{\mu \nu}}\right]\varepsilon^{\beta
\delta}_{a}\varepsilon_{\delta \gamma
\psi}+\left[\frac{f_{\mathcal{T}}}{32 \pi
}\right.\\\nonumber&\left.\times
\frac{\mathcal{T}^{\gamma(m)}_{\delta}}{f_{R}}+\frac{f_{\mathcal{L}_{m}}}{64
\pi f_{R}} \mathcal{T}^{\gamma(m)}_{\delta}-\frac{\Box f_{R}}{64 \pi
f_{R}}+\frac{\nabla^{\gamma}\nabla_{\delta}f_{R}}{64 \pi
f_{R}}+\frac{f}{128 \pi f_{R}}-\frac{R f_{R}}{128 \pi
f_{R}}-\frac{f_{\mathcal{T}}\mathcal{L}_{m}}{32 \pi
f_{R}}-\frac{f_{\mathcal{L}_{m}}}{64
\pi}\right.\\\nonumber&\left.\times \frac{\mathcal{L}_{m}}{f_{R}}
+\frac{f_{\mathcal{T}}g^{\mu \nu}}{32 \pi
f_{R}}\frac{\partial^{2}\mathcal{L}_{m}}{\partial
g^{\gamma}_{\delta}\partial g_{\mu \nu}}\right]\varepsilon^{\beta
\delta}_{a}\varepsilon_{\beta \gamma \psi},
\\\nonumber
\mathcal{N}_{a \psi}^{(B)}&=\frac{f_{\mathcal{T}}}{8 \pi
f_{R}}\mathcal{T}_{a \psi}^{(m)}+\frac{f_{\mathcal{L}_{m}}}{16 \pi
f_{R}}\mathcal{T}_{a \psi}^{(m)}-\frac{g_{a \psi}}{8 \pi f_{R}}\Box
f_{R}+\frac{1}{16 \pi f_{R}}\nabla_{a}\nabla_{\psi}f_{R}+\frac{f}{16
\pi f_{R}}g_{a \psi}-\\\nonumber&\frac{R g_{a \psi}}{16 \pi
}-\frac{f_{\mathcal{T}}\mathcal{L}_{m}}{4 \pi f_{R}}g_{a
\psi}-\frac{f_{\mathcal{L}_{m}}\mathcal{L}_{m}}{8 \pi f_{R}}g_{a
\psi}+\frac{f_{\mathcal{T}}g^{\mu \nu}g_{a \alpha}}{8 \pi f_{R}}
\frac{\partial^{2}\mathcal{L}_{m}}{\partial
g^{\alpha}_{\psi}\partial g_{\mu \nu}}-\frac{u_{\psi}u^{\delta}}{8
\pi f_{R}}f_{\mathcal{T}}\mathcal{T}^{(m)}_{a
\delta}-\frac{u_{\psi}u^{\delta}}{16 \pi
f_{R}}f_{\mathcal{L}_{m}}\\\nonumber&\times \mathcal{T}^{(m)}_{a
\delta} +\frac{u_{\psi}u^{\delta}}{16 \pi f_{R}}g_{a \delta}\Box
f_{R}-\frac{u_{\psi}u^{\delta}\nabla_{a}\nabla_{\delta}}{16 \pi
f_{R}}f_{R}-\frac{u_{\psi}u^{\delta}}{64 \pi f_{R}}g_{a
\delta}+\frac{u_{\psi}u^{\delta}R f_{R}}{32 \pi f_{R}}g_{a
\delta}+\frac{u_{\psi}u^{\delta}}{8 \pi
f_{R}}f_{\mathcal{T}}\mathcal{L}_{m}\\\nonumber&\times g_{a
\delta}+\frac{u_{\psi}u^{\delta}}{16 \pi
f_{R}}f_{\mathcal{L}_{m}}\mathcal{L}_{m}g_{a \delta}
-\frac{u_{\psi}u^{\delta}}{8 \pi f_{R}}f_{\mathcal{T}}g^{\mu
\nu}g_{a \alpha} \frac{\partial ^{2}\mathcal{L}_{m}}{\partial
g^{\alpha}_{\delta}\partial g_{\mu \nu}}-\frac{u_{\gamma}u_{a}}{8
\pi
f_{R}}f_{\mathcal{T}}\mathcal{T}^{\gamma(m)}_{\psi}-\frac{u_{\gamma}u_{a}}{16
\pi f_{R}}f_{\mathcal{L}_{m}}\mathcal{T}^{\gamma(m)}_{\psi}
\\\nonumber&+\frac{u_{\psi}u_{a}}{16 \pi f_{R}}\Box f_{R}-\frac{u_{\gamma}u_{a}}{16 \pi f_{R}}\nabla^{\gamma}\nabla_{\psi}f_{R}-\frac{f u_{\psi}u_{a}}{32 \pi f_{R}}+\frac{ u_{\psi}u_{a}}{32 \pi f_{R} }R f_{R}+\frac{ u_{\psi}u_{a}}{8 \pi f_{R}}f_{\mathcal{T}}\mathcal{L}_{m}+\frac{u_{a}u_{\psi}}{16 \pi f_{R}}f_{\mathcal{L}_{m}}\mathcal{L}_{m}\\\nonumber&-\frac{u_{\gamma}u_{a}}{8 \pi f_{R}}f_{\mathcal{T}}g^{\mu \nu}\frac{\partial^{2}\mathcal{L}_{m} }{\partial g^{\gamma}_{\psi}\partial g_{\mu \nu}}+\frac{u_{\gamma}u^{\delta}f_{\mathcal{T}}}{16 \pi f_{R}}g_{a\psi}\mathcal{T}^{\gamma(m)}_{\delta}+\frac{u_{\gamma}u^{\delta}g_{a\psi}}{32 \pi f_{R}}f_{\mathcal{L}_{m}}\mathcal{T}^{\gamma(m)}_{\delta}+\frac{g_{a \psi}u_{\gamma}u^{\delta}}{16 \pi f_{R}}\nabla^{\gamma}\nabla_{\delta}f_{R}\\\nonumber&+\frac{g_{a \psi}u_{\gamma}u^{\delta}f_{\mathcal{T}}}{16 \pi f_{R}} \frac{\partial^{2}\mathcal{L}_{m}}{\partial g^{\gamma}_{\delta}\partial g_{\mu \nu}}g^{\mu
\nu},
\\\nonumber
N_{a
\psi}^{(C)}&=-\frac{f_{\mathcal{T}}u^{\delta}\varepsilon_{\alpha a
\psi}}{32 \pi
f_{R}}\mathcal{T}^{\alpha(m)}_{\delta}-\frac{f_{\mathcal{\mathcal{L}}_{m}}u^{\delta}\varepsilon_{\alpha
a \psi}}{64 \pi
f_{R}}\mathcal{T}^{\alpha(m)}_{\delta}-\frac{u^{\delta}\varepsilon_{\alpha
a \psi}}{32 \pi
f_{R}}\nabla^{\alpha}\nabla_{\delta}f_{R}+\frac{u^{\delta}\varepsilon_{a
\gamma \psi}}{16 \pi
f_{R}}f_{\mathcal{T}}\mathcal{T}^{\gamma(m)}_{\delta}\\\nonumber&+\frac{u^{\delta}\varepsilon_{a
\gamma \psi}}{32 \pi
f_{R}}f_{\mathcal{L}_{m}}\mathcal{T}^{\gamma(m)}_{\delta}+\frac{u^{\delta}\varepsilon_{a
\gamma \psi}}{32 \pi
f_{R}}\nabla^{\gamma}\nabla_{\delta}f_{R}+\frac{f_{\mathcal{T}}\varepsilon_{a
\gamma \psi}}{16 \pi
f_{R}}\frac{\partial^{2}\mathcal{L}_{m}}{\partial
g^{\gamma}_{\delta}\partial g_{\mu \nu}}g^{\mu
\nu}u^{\delta}-\frac{u^{\delta}\varepsilon_{a \alpha \psi}g^{\mu
\nu}}{16 \pi f_{R}}\frac{\partial^{2}\mathcal{L}_{m}}{\partial
g^{\alpha}_{\delta}\partial g_{\mu \nu}},
\\\nonumber
\mathcal{C}&=-\frac{f_{\mathcal{T}}\mathcal{T}^{\alpha(m)}_{\delta}}{64
\pi f_{R}}\varepsilon^{\beta\delta\epsilon}\varepsilon_{\alpha \beta
\epsilon}+\frac{\Box f_{R}}{64 \pi f_{R}}\varepsilon^{\beta \delta
\epsilon}\varepsilon_{\alpha \beta
\epsilon}-\frac{\nabla^{\alpha}\nabla_{\delta}f_{R}}{64 \pi
f_{R}}\varepsilon^{\beta \delta
\epsilon}\varepsilon_{\alpha\beta\epsilon}-\frac{f}{128 \pi
f_{R}}\varepsilon^{\beta \delta \epsilon} \varepsilon_{\alpha \beta
\epsilon}\\\nonumber&-\frac{f_{\mathcal{T}}g^{\mu \nu}}{32 \pi
f_{R}}\frac{\partial^{2}\mathcal{L}_{m}}{\partial
g^{\alpha}_{\delta}\partial g_{\mu
\nu}}\varepsilon^{\beta\delta\epsilon}\varepsilon_{\alpha\beta\epsilon}-\frac{f_{\mathcal{T}}\mathcal{T}^{\gamma(m)}_{\beta}}{64
\pi
f_{R}}\varepsilon^{\beta\delta\epsilon}\varepsilon_{\delta\gamma\epsilon}-\frac{f_{\mathcal{L}_{m}}\mathcal{T}^{\gamma(m)}_{\beta}}{128
\pi
f_{R}}\varepsilon^{\beta\delta\epsilon}\varepsilon_{\delta\gamma\epsilon}-\frac{\nabla^{\gamma}\nabla_{\beta}f_{R}}{64
\pi
f_{R}}\varepsilon^{\beta\delta\epsilon}\varepsilon_{\delta\gamma\epsilon}\\\nonumber&
-\frac{f_{\mathcal{T}}g^{\mu \nu}}{32 \pi
f_{R}}\frac{\partial^{2}\mathcal{L}_{m}}{\partial
g^{\gamma}_{\beta}\partial g_{\mu \nu}}\varepsilon^{\beta \delta
\epsilon}\varepsilon_{\delta \gamma \epsilon},
\\\nonumber
\mathcal{D}&=\frac{\mu f_{\mathcal{T}}}{8 \pi f_{R}}+\frac{\mu
f_{\mathcal{L}_{m}}}{16 \pi f_{R}}-\frac{\Box f_{R}}{2 \pi f_{R}}
+\frac{\nabla ^{\beta}\nabla_{\beta}f_{R}}{16 \pi f_{R}}+\frac{f}{4
\pi f_{R}}-\frac{R f_{R}}{4 \pi
f_{R}}-\frac{f_{\mathcal{T}}\mathcal{L}_{m}}{\pi
f_{R}}-\frac{\mathcal{L}_{m}}{2 \pi}+\frac{f_{\mathcal{T}}g^{\mu
\nu}}{8 \pi f_{R}}\\\nonumber&\times
\frac{g^{\beta}_{\alpha}\partial^{2}\mathcal{L}_{m}}{\partial
g^{\alpha}_{\beta}\partial g_{\mu
\nu}}-\frac{u^{\epsilon}u^{\delta}f_{\mathcal{T}}}{8 \pi
f_{R}}\mathcal{T}^{(m)}_{\epsilon
\delta}-\frac{u^{\epsilon}u^{\delta}f_{\mathcal{L}_{m}}}{16 \pi
f_{R}}\mathcal{T}^{(m)}_{\epsilon
\delta}+\frac{u^{\epsilon}u^{\delta}}{16 \pi f_{R}}g_{\epsilon
\delta}\Box f_{R}-\frac{u^{\epsilon}u^{\delta}}{16 \pi
f_{R}}\nabla_{\epsilon}\nabla_{\delta}f_{R}-\frac{u^{\epsilon}}{64
\pi}\\\nonumber&\times \frac{u^{\delta}}{ f_{R}}g_{\epsilon \delta}
+\frac{u_{\beta}u^{\delta}\delta^{\beta}_{\delta}R f_{R}}{32 \pi
f_{R}}+\frac{u_{\beta}u^{\delta}\delta^{\beta}_{\delta}}{8 \pi
f_{R}}\mathcal{L}_{m}f_{\mathcal{T}}+\frac{u^{\epsilon}u^{\delta}g_{\epsilon
\delta}}{16 \pi
f_{R}}\mathcal{L}_{m}f_{{\mathcal{L}}_{m}}-\frac{u^{\epsilon}u^{\delta}}{8
\pi f_{R}}\frac{\partial^{2}\mathcal{L}_{m}}{\partial
g^{\alpha}_{\delta}\partial g_{\mu \nu}}g^{\mu \nu}g_{\epsilon
\alpha}f_{\mathcal{T}}\\\nonumber&-\frac{u_{\gamma}u^{\beta}\mathcal{T}^{\gamma
(m)}_{\beta}}{8 \pi f_{R}}f_{\mathcal{T}}
-\frac{u_{\gamma}u^{\beta}\mathcal{T}^{\gamma (m)}_{\beta}}{16 \pi
f_{R}}f_{\mathcal{L}_{m}}+\frac{\Box f_{R}}{16 \pi f_{R}}
-\frac{u_{\gamma}u^{\beta}}{16 \pi
f_{R}}\nabla^{\gamma}\nabla_{\beta}f_{R}-\frac{f}{32 \pi
f_{R}}+\frac{R f_{R}}{32 \pi f_{R}}
+\\\nonumber&\frac{f_{\mathcal{T}}}{8 \pi}
\frac{\mathcal{L}_{m}}{f_{R}}+\frac{f_{\mathcal{L}_{m}}\mathcal{L}_{m}}{16
\pi f_{R}}-\frac{u_{\gamma}u^{\beta}}{8 \pi
f_{R}}\frac{\partial^{2}\mathcal{L}_{m}}{\partial
g^{\gamma}_{\beta}\partial g_{\mu \nu}}g^{\mu
\nu}f_{\mathcal{T}}+\frac{u_{\gamma}u^{\delta}\mathcal{T}^{\gamma
(m)}_{\delta}}{4 \pi f_{R}}f_{\mathcal{T}}
+\frac{u_{\gamma}u^{\delta}\mathcal{T}^{\gamma (m)}_{\delta}}{8 \pi
f_{R}}f_{\mathcal{L}_{m}}+\frac{u_{\gamma}u^{\delta}}{4 \pi
f_{R}}\\\nonumber& \times \nabla^{\gamma}\nabla_{\delta}
f_{R}+\frac{u_{\gamma}u^{\delta}f_{\mathcal{T}}}{4 \pi f_{R}}
\frac{\partial^{2}\mathcal{L}_{m}}{\partial
g^{\gamma}_{\delta}\partial g_{\mu \nu}}g^{\mu \nu},
\\\nonumber
D_{0}&=-\frac{e^{2 \sigma}P_{r}f_{\mathcal{T}}}{8 \pi
f_{R}}-\frac{e^{2\sigma}P_{r}f_{\mathcal{L}_{m}}}{16 \pi
f_{R}}+\frac{e^{\sigma}\varrho ^{'}}{16 \pi
f_{R}}f_{R}^{'}+\frac{e^{\sigma}f_{R}^{''}}{8 \pi
f_{R}}-\frac{e^{\sigma}\sigma^{'}}{16 \pi
f_{R}}f_{R}^{'}+\frac{e^{\sigma}f_{R}^{'}}{4 \pi r
f_{R}}-\frac{e^{\sigma}f_{R}^{''}}{8 \pi
f_{R}}\\\nonumber&+\frac{e^{\sigma}\sigma^{'}f_{R}^{'}}{16 \pi
f_{R}}+\frac{f e^{2\sigma}}{16 \pi f_{R}}-\frac{R
f_{R}e^{2\sigma}}{16 \pi
f_{R}}-\frac{e^{2\sigma}f_{\mathcal{T}}\mathcal{L}_{m}}{4 \pi
f_{R}}-\frac{e^{2\sigma}f_{\mathcal{L}_{m}}\mathcal{L}_{m}}{8 \pi
f_{R}}-\frac{1}{8
\pi}\left[\frac{1}{r^{2}}-\left(\frac{\varrho^{'}}{r}+\frac{1}{r^{2}}\right)\right.\\\nonumber&\left.\times
e^{-\sigma}\right],
\\\nonumber
\chi&=\frac{e^{2 \sigma}P_{r}f_{\mathcal{T}}}{24 \pi
f_{R}}+\frac{e^{2\sigma}P_{r}f_{\mathcal{L}_{m}}}{48 \pi
f_{R}}-\frac{e^{\sigma}\varrho ^{'}f_{R}^{'}}{48 \pi
f_{R}}-\frac{e^{\sigma}f_{R}^{''}}{24 \pi
f_{R}}+\frac{e^{\sigma}\sigma^{'}}{48 \pi
f_{R}}f_{R}^{'}-\frac{e^{\sigma}f_{R}^{'}}{12 \pi r
f_{R}}+\frac{e^{\sigma}f_{R}^{''}}{24 \pi
f_{R}}\\\nonumber&-\frac{e^{\sigma}\sigma^{'}f_{R}^{'}}{48 \pi
f_{R}}-\frac{f e^{2\sigma}}{48 \pi f_{R}}+\frac{R
f_{R}e^{2\sigma}}{48 \pi
f_{R}}+\frac{e^{2\sigma}f_{\mathcal{T}}\mathcal{L}_{m}}{12 \pi
f_{R}}+\frac{e^{2\sigma}f_{\mathcal{L}_{m}}\mathcal{L}_{m}}{24 \pi
f_{R}} -\frac{f_{\mathcal{T}}P_{\bot}}{12 \pi
f_{R}}-\frac{f_{\mathcal{L}_{m}}P_{\bot}}{24 \pi f_{R}}
+\\\nonumber&\frac{\varrho^{'}e^{-\sigma}f_{R}^{'}}{24 \pi
f_{R}}+\frac{e^{-\sigma}f_{R}^{''}}{12 \pi
f_{R}}-\frac{e^{-\sigma}\sigma^{'}}{24 \pi
f_{R}}f_{R}^{'}+\frac{e^{-\sigma}}{6 r \pi f_{R}}f_{R}^{'} -\frac{
e^{-\sigma}f_{R}^{'}}{12 \pi r f_{R}}+\frac{f}{24 \pi f_{R}}-\frac{R
f_{R}}{24 \pi f_{R}}-\frac{\mathcal{L}_{m}}{f_{R}}\\\nonumber&
\times\frac{f_{\mathcal{T}}}{6 \pi
}-\frac{\mathcal{L}_{m}f_{\mathcal{L}_{m}}}{12 \pi
f_{R}}+\frac{\mathcal{L}_{a\psi}}{4 \pi}+\frac{1}{3}
\left(P_{r}+2P_{\bot}\right).
\end{align}

\end{document}